\documentstyle[]{l-aa}  

\thesaurus{03      		 
              (02.18.7; 	 
	      11.01.2;		 
   	      11.03.3;		 
	      11.09.4)}		 

\begin{document}
 
\title{The ionization mechanism of the extended gas in high redshift radio 
galaxies: shocks or AGN photoionization? }

\author{
M. Villar-Mart\'\i n,
\inst{1}
C. Tadhunter,
\inst{1}
N. Clark
\inst{1}
}
 
\offprints{M.Villar-Martin, Dept. of Physics, University of
Sheffield, Sheffield S3~7RH, UK}

\institute{$^{1}$\,Dept. of Physics, University of
Sheffield, Sheffield S3~7RH, UK\\}

\date{ }
 
\maketitle

\markboth{Ionization mechanism of the EELR in  high redshift radio galaxies.}{}
 
\begin{abstract}

We have compared the UV line ratios of a sample of very
high redshift radio galaxies (HZRG, $z > 1.7$)
with shock  and active galactic nuclei (AGN) photoioni\-zation  models, with
the goal of determining the balance between jet-induced shocks and AGN illumination in the extended emission line regions (EELR).
We find that the UV line ratios cannot be explained in terms
of photoioni\-zation of solar abundance gas by the classical
power law of index $\alpha=$-1.5, which successfully 
reproduces the general trends
defined by the optical line ratios of low redshift radio galaxies. Pure shock
models also provide a poor fit to the data. However, photoionization by a
power law of index -1.0 provides an excellent fit to the UV line ratios. 
This suggests that the ionizing continuum spectral shape may
depend on radio luminosity and/or redshift, such that it
 becomes  harder as the radio power and/or redshift increase.  However, an alternative
possibility is that we are seeing the first signs of
chemical evolution in these objects, since a power-law of 
index -1.5 with low metallicity also provides
a very good fit to the data. 

For the high ionization conditions found in the the HZRG, 
we show that the power-law photoionization mo\-dels provide a better fit to the
data than
the shock models. However, such is
the complexity of the shock models that we cannot rule out the possibility
that a different combination of input parameters can reproduce the observed
spectra.  
	
	We further show that the UV line ratios provide
a sensitive test of the ionization mechanism for the lower ioni\-zation conditions
prevalent in some low redshift jet-cloud interaction candidates. For high 
ionization parameter this discrimination is difficult due to the overlap of shock and power law photoionization models.

\end{abstract}

\section{Introduction}

The character of the emission line spectra of powerful 
radio galaxies is strongly determined
by the ionization mechanism of the gas. For most 
low redshift ($z<$0.1) radio galaxies photoionization by a 
central AGN describes rather well the general
properties of the optical emission line spectra ({\em e.g.} Robinson et al. 1987).
However, when the first high redshift radio galaxies (HZRG, $z>1.7$) 
were discovered
the situation was not so clear. These objects showed very blue 
colours and strong Ly$\alpha$ emission --- properties expected 
for galaxies in the process of formation ({\em e.g.} Spinrad {\em et al.} 1985, McCarthy {\em et al.} 1987). It was proposed that some of these  HZRG, like
3C326.1, were 
very young galaxies in which the strong Ly$\alpha$ emission was powered
by the young blue stars (McCarthy {\em et al.} 1987).
 
However,  detailed analysis of the spectra  showed that  young stars
could not explain the general properties of the UV emission lines of HZRG;
in particular,
the large emission line
equivalent widths and the existence of highly ionized species, like CIV and 
HeII ({\em e.g.} McCarthy {\em et al.} 1990).

In the context of the unified schemes for powerful radio galaxies
({\em e.g.} Barthel 1989), we would expect these galaxies 
to harbour  powerful AGN. This,
together with the fact that AGN photoionization  succeeded in
explai\-ning the optical emission line ratios of
low redshift radio gala\-xies, suggested that the same mechanism dominates
the ionization processes in HZRG (McCarthy 1993). In consequence,
most attempts to explain  the UV emission line spectra of HZRG have 
invoked
pure AGN photoionization.

	However, both imaging and spectroscopy of HZRG show that  strong interactions are taking place between the advancing radio 
jet  and the ISM of the host galaxy (section 2). Such interactions
will generate powerful shocks which  will disturb the morphology, kinematics 
and physical conditions (density, temperature, pressure)
 of the gas and potentially modify its ionization state. 
Currently a key issue in the study of
these objects is the relative importance of jet-induced shocks 
and AGN illumination:
do shocks dominate the emission line processes? Or is  AGN photoionization dominant
with the influence of
the shocks
mainly manifested  in kinematic and morphological disturbances?

	We have addressed this problem by studying the UV emission line spectrum
of a sample of very HZRG ($z>$1.7), comparing their line ratios with both shock and AGN photoionization models. We base our study on the UV lines because
most of the information we have about HZRG  is derived from 
studies of the UV emission
line spectra.  At such high redshifts the main optical diagnostic
emission lines are shifted into the infrared, and most
of the exis\-ting IR spectra for the HZRG are of poor quality. 

In contrast, at low redshifts we have the opposite pro\-blem: the
optical diagnostic line ratios are well measured, but the UV spectra
are of low quality. For the low redshift objects the optical
line ratios have proved inefficient at distinguishing the 
ionization mechanism unambiguously, although it has been
suggested  that the UV line
ratios might provide a stronger discriminant (Sutherland {\em et al.} 1993).  
Therefore, the interest in
understanding the emission of the UV lines can be extended to low redshift
objects. We propose to develop a diagnostic method to discriminate
between shock and AGN ionization in radio galaxies at all redshifts.

	We review in section 2 the observational evidence for AGN illumination
and shocks in powerful radio galaxies. The data sample is described in section
3 and the diagnostic diagrams in section 4. Sections 5 and 6 present the models
and the comparison  with the data: section 5 concentrates
on the effects of a) varying   the shape
of the AGN  continuum and b) changing the excitation and ionization mechanism (shocks); section 6 analyzes the effects that different physical
conditions in the extended gas can produce in the observed UV spectra.
In section 7 we extend our diagnostic method to low redshift radio galaxies.
Section 8 includes summary and  conclusions.

\section{Shock {\it vs.} AGN photoionization in HZRG ($z>$1)}
 
\subsection{Evidence of AGN illumination in HZRG.}

 In the popular anisotropic illumination model ({\em e.g.} Fosbury 1989), it is 
postulated that quasars hidden in the cores of powerful radio galaxies 
illuminate the ambient ISM with intense cones of UV/X-ray radiation, with the 
radiation photoionizing the extended gas within the cones, leading to  
line emission. This model appears to be consistent with the observed properties 
of the majority of {\em low-redshift} radio galaxies. Most importantly, the 
line ratios measured in the nuclear regions and EELR
of radio galaxies, and the 
trends in these line ratios, generally agree well with photoionization models 
(Robinson {\em et al.} 1987; Binette {\em et al.} 1996). The (weak) alignment 
of the EELR with the radio axes (Baum \& Heckman 1989) and
rela\-tively undisturbed kinematics (Tadhunter et al. 1989) are also consistent 
with anisotropic illumination by the  broad radiation cones predicted
by the unified schemes. Although there are no radio galaxy EELR which 
show the clear cone-like morphologies seen in some Seyfert galaxies, this 
is likely to be due to relatively sparse and inhomogeneous  distribution 
in the early-type host galaxies (Tadhunter 1990). 

Anisotropic AGN-photoionization of the ambient ISM is also a viable model for 
 {\em high-redshift} radio galaxies. 
The detection of  scattered light from a hidden quasar 
(and also broad line components) in the polarized flux of seve\-ral HZRG is  strong evidence
for the existence of luminous quasars which illuminate the ISM ({\em e.g.} Cimatti {\em et al.} 1996, Dey {\em et al.} 1996). Moreover, 
the presence of large diffuse halos of
ionized gas in some HZRG, which extend far beyond the radio structures
strongly suggests the existence of a quiescent ISM ionized by the central AGN
({\it eg.} van Ojik 1995).

	McCarthy (1993) constructed a composite radio galaxy spectrum from
observations of galaxies with 0.1$<z<$3. Photoionization calculations
reproduce the radio galaxy spectrum rather well and this was presented as
an argument in favour of the AGN photoionization as the main ionization
mechanism in radio galaxies. However, the composite was built from
the spectra of very different objects covering a wide range in redshift, and
it is dominated by one
or two of the most highly ionized objects. 
Thus the comparison of this composite spectrum with the models does not 
resolve the issue of the ionization mechanism for the general population
of HZRG.

	Villar-Mart\'\i n et al. (hereafter VMBF96) concluded that 
photoionization by the AGN can explain the positions of a large sample
of HZRG in the  CIV$\lambda$1550/Ly$\alpha$ {\it vs.} CIV$\lambda$1550/CIII] $\lambda$1909 diagnostic diagram. The sequence defined by the data can
be parametrized in terms of the so-called ionization parameter, that is, the ratio of
the density of ionizing photons impinging on the slab to the density of the outermost
gas layer of the slab:

	$$ U = \frac{1}{cn_H} \int_{\nu_0}^{\infty} {\frac{\phi_{\nu}}{h\nu} d\nu} ~~~ [1]$$

where $c$ is the speed of light, $n_H$ is the density of the gas in the front layer and $\nu_0$ is the Lyman limit frequency.
$\phi_{\nu}$ is the monochromatic ionizing energy flux  impinging
on the slab. By varying the ionization parameter it is possible to produce the variety observed in the UV line ratios of HZRG. A similar result
is obtained at low $z$: the sequences defined by the optical line ratios
of powerful radio galaxies are  explained in terms of a sequence
defined by $U$ (Robinson et al. 1987).

\subsection{Evidence of shocks in HZRG.}

A possible alternative ionization 
mechanism in these sources is the ionization by fast shocks produced by violent 
interactions between the advancing radio jet and the ambient gas: there is 
already clear evidence for such jet-cloud
interactions in HZRG. Firstly, whereas the gas 
kinematics in most nearby radio galaxies are consistent with gravitational 
motions (Tadhunter {\em et al.} 1989; Baum {\em et al.} 1992), extreme 
non-gravitational motions are observed along the radio axes in the majority of 
HZRG (van Ojik 1995; McCarthy {\em et al.} 1996). Secondly, the extended emission line regions (EELR) are often not only aligned with the radio axis, but closely correlated 
in detail with the radio emission (Chambers {\em et al.} 1990; Miley {\em et 
al.} 1992; Rigler {\em et al.} 1992; van Ojik 1995). Even when there are no 
direct radio/optical associations, the degree of collimation seen in the narrow 
jet-like EELR along the radio axes of sources like 3C~368 and 3C~324 (Longair {\em et al.} 1995) is 
difficult to explain without invoking interactions between the line-emitting 
gas and the radio jets. The radio-optical asymmetries (McCarthy {\em et al.} 
1991), the relationship between optical structure and radio size (Best {\em et 
al.} 1996), and the fact that the extent of the line-emitting gas is almost 
always smaller than that of the associated radio source (van Ojik 1995), provide
further evidence for a close association between the radio plasma
and the warm emission line gas. 

Recently, we have made a detailed study of the EELR in a sample of low--intermediate redshift 
radio galaxies which show clear morphological evidence 
for jet--cloud interactions (Tadhunter {\em et al.} 
1994; Clark \& Tadhunter 1996; Clark 1996; Clark {\em et al.} 1996). This
study provides  clear evidence that jet-induced shocks determine the 
distribution, kinematics and physical conditions  
of the EELR along the radio
axes of the objects in the sample. There is also evidence that shocks have an ionizing 
effect in these sources: in particular, the high  temperatures indicated by the
the [OIII]4363/(5007+4959) ratio, and the low 
HeII(4686)/H$\beta$ ratio measured in the extended gas, are more consistent 
with shock-ionization than AGN-photoionization. Even complex multi-phase 
photoionization models such as those presented recently by Binette {\em et al.} 
(1996) and Simpson \& Ward (1996) cannot  reproduce  the measured 
values of these two line ratios.

With the above discussion in mind, it is imperative that the UV line ratios of 
HZRG be compared in detail with the predictions of both shock-ionization and 
AGN-photoionization models.

\section{The data}

We have constructed a sample which comprises  21 narrow line
radio galaxies at high~z (most of them have $z>$2) for
which the UV CIV$\lambda$1550, HeII$\lambda$1640
and CIII]$\lambda$1909 emission lines
have been measured.   We have excluded those objects which show evidence
of a BLR. In Table~1, we list the object names, the line ratios of
interest to us here, the redshift and the reference to the
observations. Most of the data are taken from the
recent thesis by van Ojik (1995) which includes objects selected on
the basis of a very steep radio spectrum.  The other objects have been
selected from the literature with the criteria of having high redshifts, 
classified as  narrow line radio galaxies and with measured fluxes
for the three UV lines mentioned above.
      
The line
measurements refer to the integrated emission from the object
collected with a long slit aligned with the radio axis. We know that
at this redshift, the line luminosities are generally dominated by the 
extranuclear emission and we can be confident that the general properties
of the emitting gas 
inferred from the spectra describe the  
extended emitting regions, rather than the emission line regions close
to the nuclei.

\begin{table*}
\centering
\caption{Observed UV line ratios for several high~z radio galaxies with no apparent broad component }
\begin{tabular}{llllll} \hline
\hline

Name  	 & 	CIV/CIII]  & 	CIV/HeII & CIII] /HeII &  redshift  & ref. \\ \hline \hline
Average HZRG &  2.05 &  1.14 &  0.56 &  & McCarthy 1993 \\ \hline 
F10214+4724 &	3.68  & 1.43	&   0.39     &  2.29 & 	Elston {\em et al.} 1994	\\ 
3C294   &    0.83  &      1.00 &  1.20   &   1.79 &  McCarthy {\em et al.} 1990 \\
3C256-3C239 &	1.90   &	    1.94  &	1.05 &  1.96  & Spinrad {\em et al.} 1985	\\
0200+015 & 	1.05	 &      1.31 &   1.25  &  2.23	& van Ojik 1995 \\ 
0211-122 & 	2.55	 &      1.81    &    0.71 &  2.34	&  ~~~~~~ " \\  
0214+183 & 	1.67	 &      1.67   &      1.00  &  2.13 	& ~~~~~~ " \\ 
0355-037  &	1.17	 &      0.73	  &  0.62 &  2.15 & ~~~~~~ " \\
0448+091 & 	0.44	  &     0.86	 &    0.86   &   2.04 & 	~~~~~~ " \\
0529-549 & 	0.22	 &      0.67 & 	   3.05   &  2.58 & 	~~~~~~ " \\
0748+134 & 	1.29	  &     1.20  &        0.93   &   2.42 & 	~~~~~~ " \\
0828+193 & 	0.95	 &      1.00       &     0.95 &  2.57 & 	~~~~~~ " \\
0943-242 & 	1.70	  &     1.44 &       0.85   & 2.92 & 	~~~~~~ " \\
1138-262 & 	0.62    &           0.62   &     1.00  &  2.16  & 	~~~~~~ " \\
1410-001 & 	1.58	 &      1.44  &      0.92  &   2.36 & 	~~~~~~ " \\
1558-003     &     2.25 &     1.59    &    0.71 & 2.53  & 	~~~~~~ " \\
2251-089 & 	2.20	 &     2.54     &   1.15 &  1.99 & 	~~~~~~ " \\
4C23.56	     &     3.40	 &     3.92   &     1.15 &   2.48 &	~~~~~~ " \\
4C26.38	 & 	3.71   &     	1.56     &   0.42 &  2.88  & 	~~~~~~ " \\
4C28.58	 & 	0.17	 &      0.19  &     0.89  &  2.89 & 	~~~~~~ " \\
4C40.36	 & 	1.05	 &    1.11  &      1.06   &  2.27 & 	~~~~~~ " \\
4C48.48  &      2.18 &     1.65  &      0.76  &  2.34 & 	~~~~~~ " \\
\hline
\hline
\end{tabular}
\end{table*}

\section{The UV line ratios: the diagnostic diagrams.}

	We have studied three diagnostic diagrams which involve the 
CIV$\lambda$1950, CIII] $\lambda$1909 and HeII$\lambda$1640 lines which usually dominate
(together with Ly$\alpha$) the UV spectrum of HZRG. 
 We compare the position in the diagrams of the data with the prediction of pure AGN  photoionization 
 and shock models. Shock model calculations
for the Ly$\alpha$ line do not exist in the literature and we have
not included  this line in our diagrams (see VMBF96 for a detailed study of the  CIV$\lambda$1950/Ly$\alpha$ {\it vs.} CIV$\lambda$1950/CIII] $\lambda$1909 diagnostic diagram for the same sample of objects studied
here and AGN photoionization models).

The data  are represented in the diagrams with hollow circles (van Ojik's data)
and solid triangles (other objects). The different models are distinguished
by different line types in the diagrams (see Figure captions). 

 Comparing data and model predictions we want to answer the questions: is it possible to
disentangle the main mechanism responsible for the ionization of the gas in HZRG using the UV line ratios? Which models agree better with the data: shocks or AGN photoionization?

\section{The influence  of the ionizing mechanism on the UV line ratios.}

\subsection{AGN photoionization models.}

 We have used the multi-purpose photoionization code MAPPINGS~I to build these
models. The version
des\-cribed in Binette et~al.  (1993a,b) is well suited for the conditions
studied here, since it includes a detailed treatment, not only of the transfer of optically thin lines (HeII$\lambda$1640  and CIII] $\lambda$1909) but also of the resonant lines, Ly$\alpha$ and CIV$\lambda$1550, which have strongly-geometry  dependent emission (VMBF96).

	The first step to predict the spectrum emitted by a gaseous region 
is to define the
 physical conditions of the gas and the shape of the ionizing continuum. At very high redshift, there are large uncertainties
in this respect. The optical lines, which have provided a great deal of 
information about the emitting gas  and the ionization
processes involved in low redshift radio galaxies, are redshifted
into the IR. As mentioned before, the optical (rest-frame) spectra obtained for very high redshift so
far  are rare. Therefore, it is not
possible to use optical line diagnostics to constrain the physical conditions
of the emitting gas in our objects.

	As a starting point, it is reasonable to assume
 that these regions are similar to the EELR in low redshift radio galaxies.
The discrepancies, if they exist, between the data and the standard models which
are successful for the low-$z$ objects will allow us to build more coherent models and therefore,
reveal information about the actual conditions of the emitting gas and the nature
of the ionizing source. Therefore, the question we will answer in this 
section is: {\it can the  models which reproduce the optical line ratios
of low redshift radio galaxies also  explain the UV line ratios of very 
high redshift radio galaxies?}

	Analysis of the optical emission lines has
 shown that typical densities in EELR 
apparently photoionized by the central AGN at low
redshift are  lower than a few hundred. Indeed, if the line emitting gas
in the EELR is in pressure equilibrium with the hot phase of the ISM,
the EELR densities implied from X-ray observations 
are of the order or a few particles per cm$^{-3}$ (Clark 1996).
 McCarthy (1993) reached a similar conclusion using a different approach
based on the Ly$\alpha$ luminosity.
Shocks, however, can raise the density 
 up to a few hundred (Clark {\em et al.} 1996). This increase in the density is not large 
enough that we need to worry about the suppression of the emission lines due
to collisional deexcitation effects. 
In our ``pure AGN'' photoionization
model,  such an increase in the  density will have equivalent effects
to a decrease in the flux of the AGN ionizing photons by the same factor. This
means that those objects where shocks have compressed the emitting gas 
-- but not ionized it --can be interpreted  in terms of a lower
ionization parameter (see eq.~1). 

 We have assumed a density of 10 cm$^{-3}$
at the illuminated face.  The behaviour of the gas is 
isobaric. Therefore, the density at every position in the cloud is adjusted
with the temperature to keep the pressure equilibrium.
The abundances are assumed to be solar unless another value is specified. The clouds are considered to be radiation bounded (except in section 6.2) and with plane
para\-llel geometry (see VMBF96 for a more detailed description).

\subsubsection{Power Law models, $\alpha$=-1.5}

	We have first investigated  whether -1.5 power-law (PL) photoionization models,  which are so successful for low-$z$ radio galaxies,  are also able to explain 
the  UV line ratios of our sample of high redshift objects.
  The results are presented in the  diagnostic diagrams in Fig.~1.
The -1.5 PL sequence is represented on the left diagrams by  solid circles  connected with a solid
line. 
	
	A quick look to the left diagrams shows that the -1.5 sequence lies
far away from the data points. However, comparing 
 the shape of the model sequence and the 
trend defined by the data,  the similarity suggests that the data sequence can be explained  in terms of the variation in the ionization parameter, as is
 observed at low redshift.

\begin{figure*}[htb]
\includegraphics{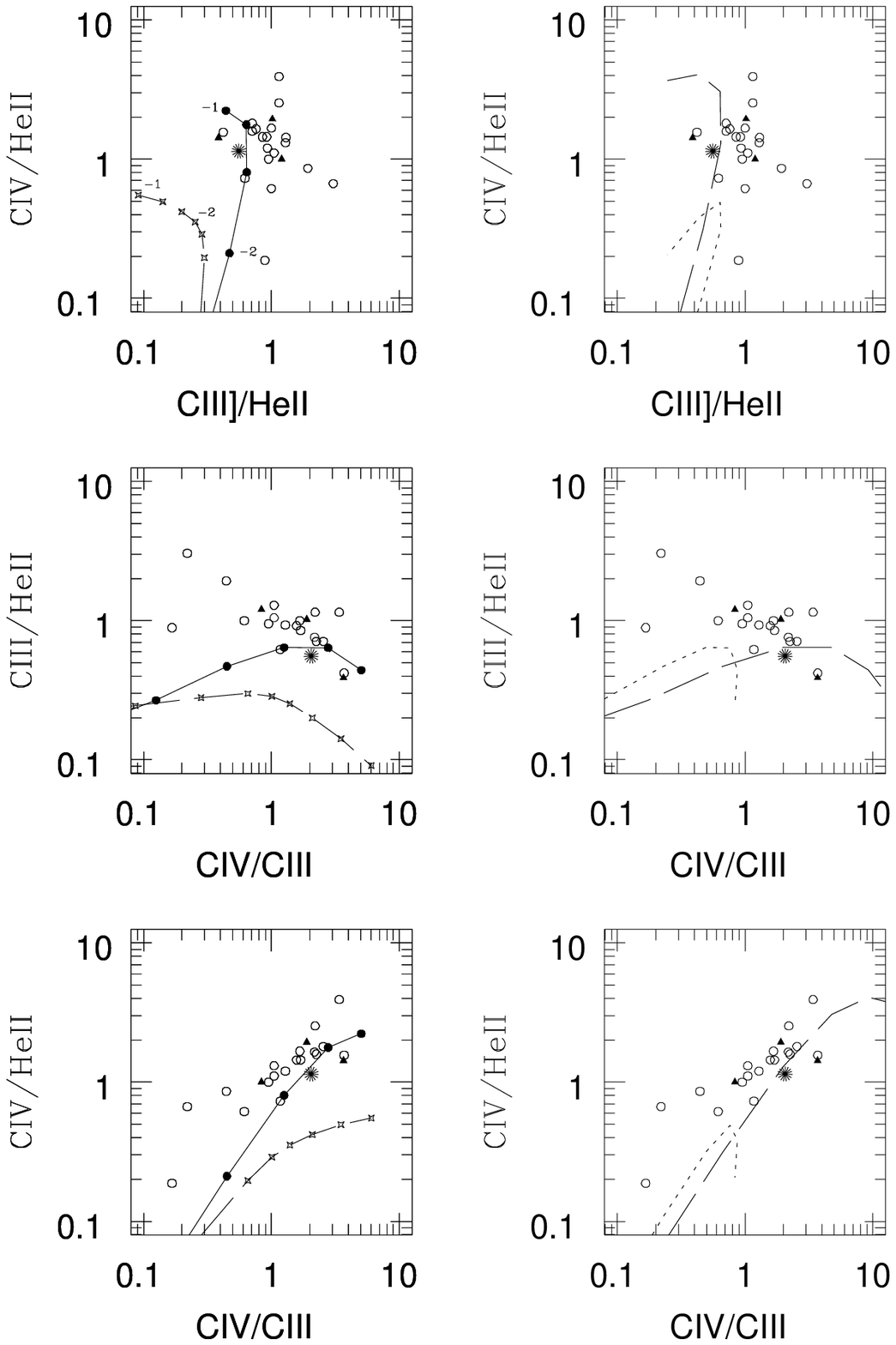}
\vspace{9in}
\caption[]{{\bf Left:} Comparison of the data sample of HZRG with  photoionization models
in which the ionizing continuum is: 1) a power law of index -1.5  (solid circles connected by a  solid line) 2) Hot black body, T$\sim$130000 K
(stars  connected by a long-dash line). The negative numbers indicate log$U$. 
Hollow circles are van Ojik's (1995) HZRG and solid triangles other HZRG 
with published UV line fluxes. The star corresponds to the average HZRG 
spectrum of McCarthy (1993).
{\bf Right:} The two new sequences use a -1.5 power law as ionizing
continuum but with geometrical effects considered. The long-dash line
describes the spectrum of the clouds seen directly from the illuminated face.
The dotted line corresponds to the opposite case, the clouds are observed from
the ``dark'' face. (Model dots have been suppressed for simplicity).}

\end{figure*}

\vspace{0.2cm}

{\it The influence of geometrical effects}

\vspace{0.2cm}

VMBF96 showed that -1.5 PL models can explain the sequence defined by the
objects in the Ly$\alpha$/CIV {\it vs.} CIV/CIII]  diagnostic diagram when geometrical effects are taken into
account. The resonant character of the CIV$\lambda$1550 and Ly$\alpha$ 
lines makes the escape of the line photons
very asymmetric, so that their emission  is strongly dependent on the distribution
of material inside (and outside, for Ly$\alpha$) the ionized cones. 
Provided that the axis of extended emission line
nebulosity is not in the plane of the sky, the spectrum we observe is emitted by a mixture of clouds observed
from different viewing angles: the clouds further from the observer are seen 
preferentially
 from the illuminated face, while the clouds closer to the observer are 
seen from the rear. In this situation resonant lines are
emitted very differently on the two sides of the nucleus; they appear stronger  
 with respect to the other lines  on the side which lies
further from the observer, and fainter on the other side.

The previous
sequences (Fig.~1, left) did not consider the geometrical effects described
above:
they  represent a situation in which  geometrical effects due to different orien\-tation angles are cancelled.

	In Fig.1 (right)  we show the same diagnostic diagrams as before, but 
geometry is considered. The single PL sequence in the left diagrams has been replaced by two new ones: the long dashed line represents models for clouds observed
directly from the illuminated face --- this des\-cribes the case in
which the spectrum of the gas on the far side of the source is dominant --- 
while the  dotted line corresponds to the opposite case in which we observe
the clouds from the rear (i.e. the clouds on the near side domi\-nate).  
Any intermediate case is described by a sequence
of models intermediate between these.  

The back perspective does not help at all to
improve the fitting to the data because the resonant CIV line intensity
is decreased relative to the other lines. For the ``front'' sequence, the line
ratios involving CIV increase slightly, but not enough to explain
the relative faintness of HeII. Moreover,  CIII]/HeII is still a problem: geometrical
effects do not affect this line ratio and cannot explain its high observed values.

	The conclusion is that  {\it photoionization by a power law of index -1.5, which
reasonably reproduces the optical line ratios of low redshift radio galaxies
 cannot explain the observed UV line ratios of HZRG}.

\subsubsection{Hot Black Body models}

	A very hot black body (T$\sim$130,000 K) provides an even better fit to the optical line ratios of low redshift radio galaxies than the -1.5 power law  usually assumed (Robinson {\em et al.} 1987). 
We have studied the predictions of these
models in the UV spectral range. The results are presented in Fig.~1 (left
pannels).
Contrary to the conclusion obtained in the optical range, such a continuum {\it cannot}
explain the position of the objects in the UV diagnostic diagrams. The 
fit is worse than the one produced by the -1.5 PL models --- the
T$\sim$130,000 K black body model falling even further
from the data points. The dispersion
 produced by geometrical effects
is similar to the -1.5 PL models. We do not present these models here for simplicity.

Therefore, the models which are successful at explaining  the low redshift optical spectra do not work
in the UV spectral range for very high redshift objects.
An alternative possibility is that the  ionizing continuum
emitted by the central AGN has a different shape at low and high redshifts. 
We investigate now the effects that a harder continuum have on the
UV line ratios.

\subsubsection{A hard ionizing continuum: Power law with index $\alpha$=-1.0}

The new models, with a power law of index $\alpha$=-1.0, are shown in the diagnostic diagrams in Fig.~2. As in Fig.~1, the diagrams on the left do not take into
account geometrical effects, which, on the contrary, are considered on the right
plots.
 The agreement
with the data is excellent. Most of the data points are
rather well described by the $\alpha$=-1.0 U-sequence.

\begin{figure*}
\includegraphics{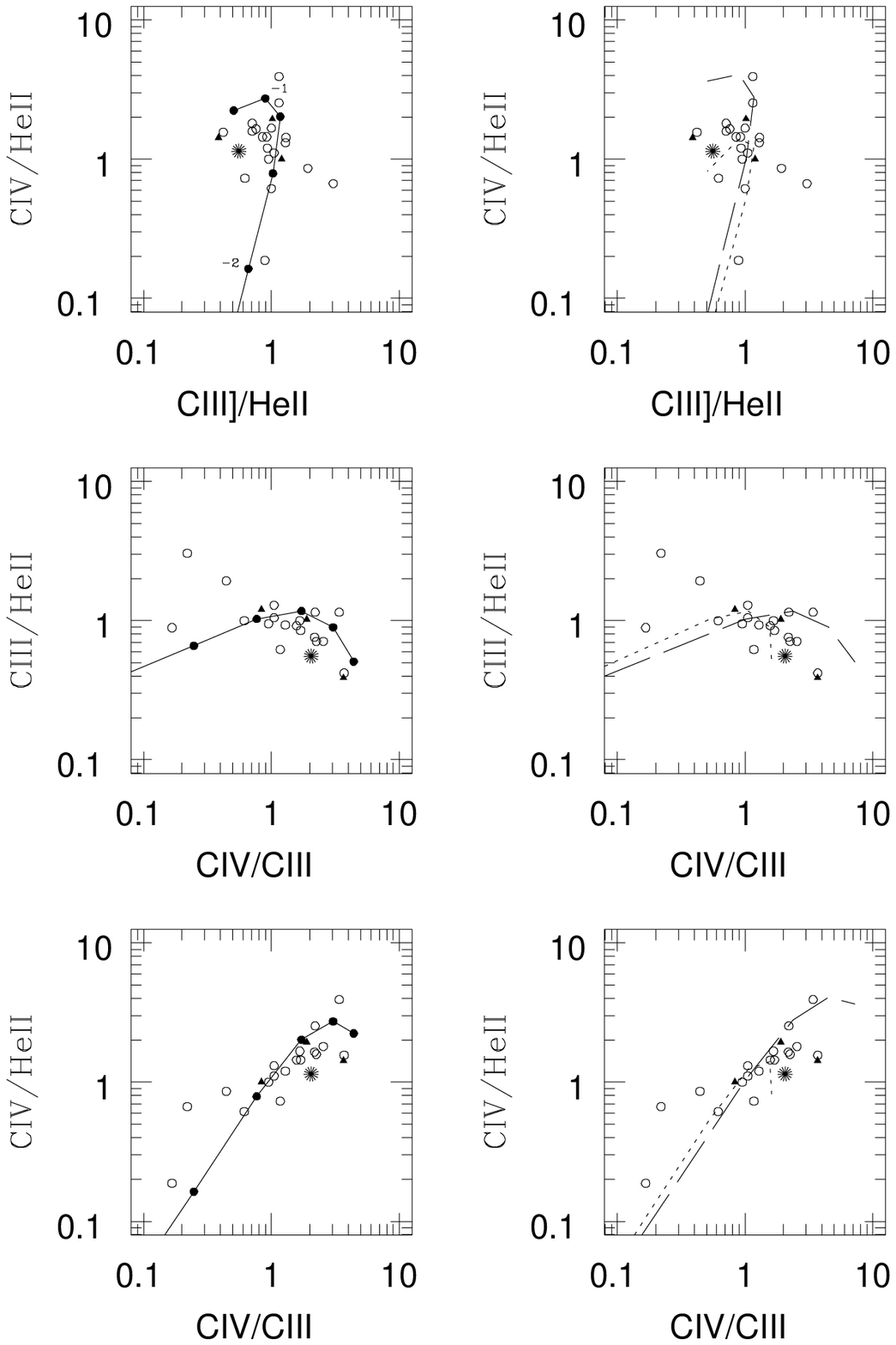}
\vspace{9in}
\caption[]{ The ionizing continuum is in the new sequences a power law
of index $\alpha=$-1.0 (symbols as in Fig.~1). The improvement to fit the observed positions is remarkable.}

\end{figure*} 

\vspace{0.2cm}

{\it The influence of geometrical effects}

\vspace{0.2cm}

When considering geometrical effects with the $\alpha$=-1.0 models (Fig.~2, right), most objects lie between the two sequences which represent the back and the front perspective, as expected. For a fixed $U$ we find that, while CIII] /HeII is the same for back and front, the dispersion
predicted by the models for different viewing angles is large  when line ratios
involving CIV are considered and the 
ioni\-zation level of the gas is high. This trend is also suggested by the data: the
models seem to envelope a very similar area to the one occupied by the data. 
Note that the -1.0 PL models can explain the
data points in the CIV/Ly$\alpha$ {\it vs} CIV/CIII] diagnostic diagram as well
as the 1.5PL mo\-dels.

 	In summary,  {\it a power law of index $\alpha$=-1.0, where the sequence is defined by the variation of
the ionization level of the gas, can explain the observed UV line ratios of
HZRG.}

If this result is confirmed,
it indicates an evolution of the spectral  shape of the ionizing 
continuum emitted by the central AGN with redshift or  luminosity (we are biased to very powerful objects). 

	 There are already some indications of a correlation between the hardness
of the AGN continuum  and redshift.
O'Brien {\em et al.} (1988) found that quasars of higher redshift show harder
UV continua. Francis (1993)  concluded that high redshift (z=2) AGNs have 
intrinsically harder mean continuum slopes ($\Delta\alpha\sim$0.8) than low redshift AGNs. He
also concludes that this is  a correlation with redshift and not
with absolute magnitude. It is important to mention that these authors study a spectral range longward (in wavelength) of the Lyman
limit and, therefore, they do not include the UV ionizing continuum of
interest to us in this paper. However, extrapolating their results
to shorter wavelengths, the ionizing continuum should also get harder
with increasing redshifts.

\subsection{The shock models}

	Another strong possibility to explain the discrepancies  between
the -1.5 PL (and hot black body) models and the observed UV line ratios is shocks.
A high velocity radiative shock can influence the emission line processes
through two different mechanisms: a) the generation of a strong local 
UV photon field in the hot post-shock zone,
which can ionize the surrounding medium both upstream and downstream; b)
line emission during  
the radiative ~cooling of gas behind the hot post-shock zone.
We have used the new shock models by Dopita and Sutherland (1995) which
take into account both effects.  
The two main parameters which
influence  the predicted spectrum  are the velocity of the shock and the magnetic
parameter defined as $B/n^{1/2}\mu$G.cm$^{-2}$, where $B$ is the pre-shock 
magnetic transverse field and $n$ is the pre-shock density. Note that we
have assumed an ionized helium recombination ratio of 
HeII$\lambda$1640/He$\lambda 4686\sim$7.7 in order to calculate
the HeII$\lambda$1640 strength from the HeII$\lambda$4686 strengths
published in Dopita \& Sutherland (1995). Our experience with
photoionization models shows that the 
value of the HeII$\lambda$1640/He$\lambda 4686$ recombination ratio
is insensitive to the physical conditions and the ionization state
of the gas.

\begin{figure*}[htb]
\includegraphics{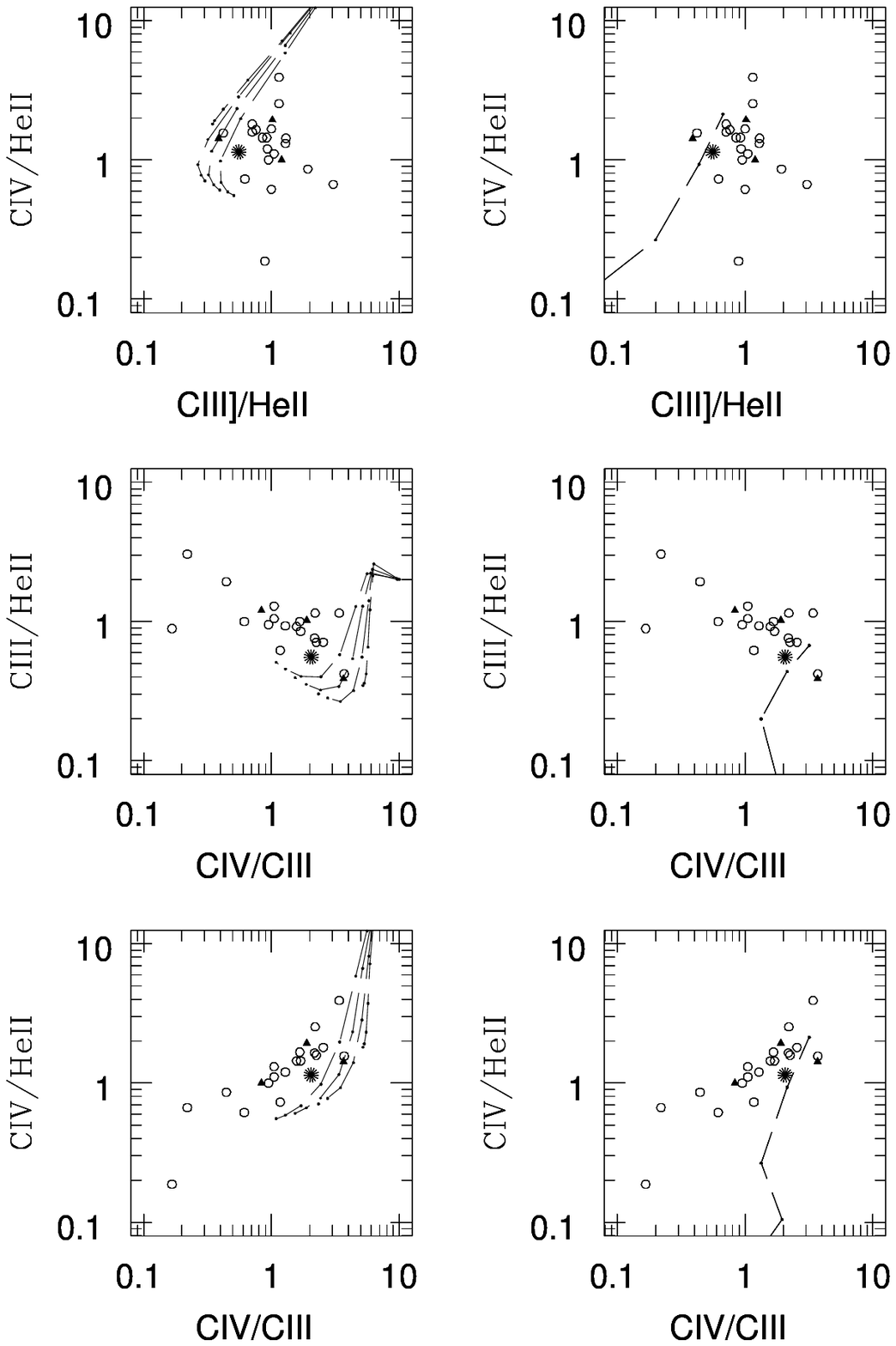}
\vspace{9in}
\caption[]{Comparison of the data sample of HZRG with the new
shock models by Dopita \& Sutherland (1995):
the diagrams on the left show the UV line ratios predicted for the
cooling radiation lines. The diagrams on the right present the models for
the precursor gas, ionized by the upstream photons emitted by the
shock. These shock models are unable to explain the observed
line ratios. For the cooling
radiation models, both velocity (150-500 km s$^{-1}$) and magnetic parameter (0-4 $\mu$G cm$^{-3/2}$) change, while for the precusor the velocity is varied (200-500 km s$^{-1}$) with a fixed magnetic parameter (1 $\mu$G cm$^{-3/2}$). }

\end{figure*}

	Shock model sequences are presented in the diagnostic diagrams in Fig.~3.
The diagrams on the left show the UV line ratios of the cooling region
(post-shock material). The velocity varies between 150 and 500 km s$^{-1}$
and the magnetic parameter between 0 and 4  $\mu$G cm$^{-3/2}$. The density
adopted is n(H)=1cm$^{-3}$.
The diagrams on the right co\-rrespond to the
precursor gas emission, that is, material which has not entered the shock but is
ionized by its UV photon flux. The velocity is varied (200-500 km s$^{-1}$) with a fixed magnetic parameter (1 $\mu$G cm$^{-3/2}$).

 The agreement between the shock and precursor mo\-dels with the data
is very poor. For the range of 
models which reproduce the observed range of CIV/HeII values, CIII]  is too
faint, both with respect CIV and HeII, so that the sequences cannot cover
the area occupied by the objects.  This can be said both for the post-shock and
precursor models. Therefore, a mixture will still produce too faint CIII].

	Another problem evident from the diagrams is the di\-fficulty
of the  models at reproducing the sequences defined by the data. In
contrast, the trends on the diagnostic diagrams are 
naturally explained in the AGN photoionization context by a sequence in
the ionization parameter $U$.  
 
	The conclusion is that {\it the published shock models cannot explain the
observed UV line ratios}. However, shock models
are complex and we cannot rule out the possibi\-lity that a combination
of parameters will be found that reproduces the observed spectra.

\section{The influence of the physical properties of the gas.}

\subsection{The metallicity}

So far, we have considered that differences in the ioni\-zing mechanism
  determine
those in the emission line spectra. The  physical conditions of the gas can also influence strongly the emission line processes. An
interesting and reasonable possibility is that the metallicity of the gas at
high redshifts is lower than solar. We have calculated models in which
 the abundances
of the heavy elements are 0.4 times the solar values. The ionizing continuum 
is the traditional -1.5 PL.

	The agreement with the data is excellent as the diagrams in Fig.~4
show. The models are represented by the solid line connected with solid
circles. Again, these mo\-dels can also explain the
data points in the CIV/Ly$\alpha$ {\it vs} CIV/CIII] diagnostic diagram.

\begin{figure*}
\includegraphics{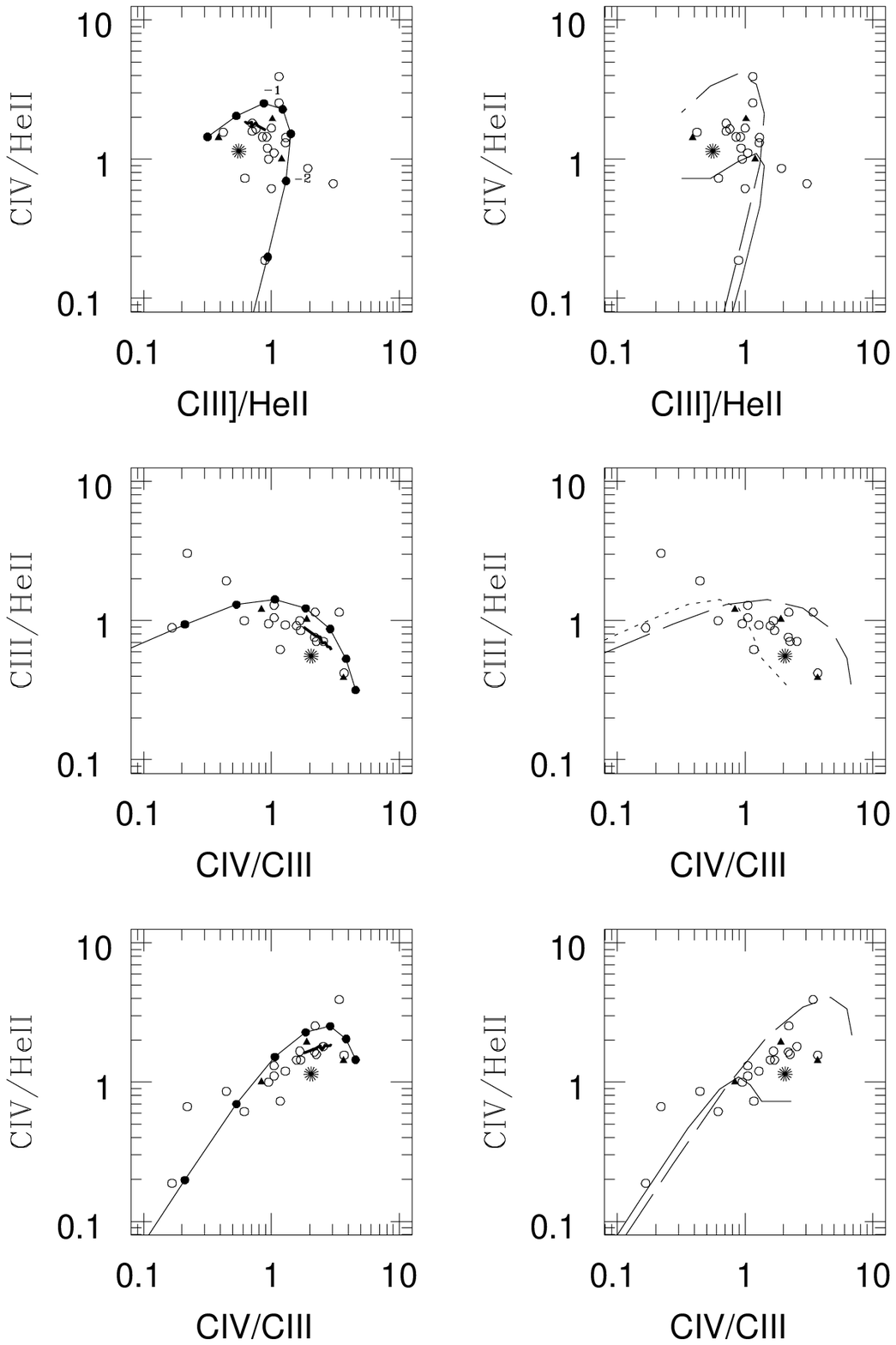}
\vspace{9in}
\caption[]{ The ionizing continuum is in the new sequences a power law
of index $\alpha=$-1.5 (symbols as in Fig.~1). In this case the abundance
of the heavy elements are decreased by a factor of 0.4 with respect the
solar values. The agreement between the models and the data is excellent.
The thick short solid line corresponds to the sequence of models proposed
by Binette et al. (1996) with a mixture of matter bounded and radiation
bounded clouds.}

\end{figure*}

	Since the CIV/CIII]  line ratio is strongly dependent
on $U$ (and not so much on metallicity in 0.4 solar
to solar range of values considered) it is interesting to compare the $U$ values required to reproduce
the CIV/CIII]  line ratio with the values derived for low redshift EELR from the optical line ratios. Most high redshift objects lie in the log$U$ range [-2,-0.5].  The ionization level of the gas is similar to that observed
in highly ionized EELR at low redshift (Robinson {\em et al.} 1987). 
However, this is not the
trend observed in low redshift jet---cloud interaction objects, where shocks are important (Clark 1996, Clark {\em et al.} 1996). In these cases the average
level of ionization of the gas is lower than that measured for the EELR of
radio galaxies where AGN photoionization dominates. If a -1.5 PL
is responsible for the ionization of
the gas, these results  suggest that we are biased towards
very powerful objects in which the gas is  highly ionized.

	If lower metallicities are confirmed at high redshift a 
harder  AGN ionizing continuum is no longer necessary and
the classical -1.5 PL is still valid.
An interesting consequence is that this would be
the first clear evidence for {\it chemical evolution with
redshift} in the host galaxies of powerful radio galaxies.

	There is other  evidence for chemical evolution with
redshift, with decreasing metallicities for younger
objects: damped Ly$\alpha$ systems  (DLA) at large redshifts ($z\sim$2-3), which
are thought
to be produced by protogalactic disks, are measured to have
a metallicity $Z\sim$1/10
$Z_{\odot}$ (Pettini {\em et al.} 1994).  Although the metallicity evolution revealed
by the DLA might not be valid for all kind of objects, 
it is suggestive and might
also happen in radio galaxies.

\subsection{Matter bounded clouds}

In spite of the good general agreement between -1.5 PL models and the observed optical
line ratios of {\it low} redshift radio galaxies, there are three problems which the classical photoionization models cannot
explain: a) too weak
high ionization lines; b) too low electronic temperatures; c) too small range in the ratio HeII$\lambda$4686/H$\beta$. Several authors have
proposed in the past that a contribution
of matter bounded clouds could help to solve some of the discre\-pancies ({\it e.g.} Viegas \& Gruenwald 1988).
In a recent paper Binette {\em et al.} (1996) develop a detailed study around this possibility. They
consider two distinct populations
of line emitting clouds: a matter bounded component and an ionization bounded component, with the ionization bounded clouds illuminated by the
ionizing spectrum escaping from the matter bounded component.
A variation in the ratio of the two components  
can explain the  sequences in the optical
line ratios on the diagnostic diagrams, and they can 
also solve the three problems
mentioned above.

We have  considered these models as a possible solution to the discrepancies
between the -1.5 PL models and the data. 
The authors present the predictions  for the  CIV/CIII] , CIV/H$\beta$ 
and HeII$\lambda$4686/H$\beta$ ratios. Asu\-mming as before,
HeII$\lambda$1640/HeII$\lambda$4686$\sim$7.7 we can compute the UV line ratios of interest to us.

	The models are represented in Fig.4 (short thick solid line). 
The authors adopt as continuum energy distribution a power law of index
-1.3  and select
a constant ionizing parameter 0.04.
The variable parameter is $A_{I/M}$, which re\-presents the collecting area
ratio of the ionization bounded clouds to radiation bounded clouds. The values
vary from 0.06 to $\sim$4.

These models are unable to explain the wide
range in UV line ratios observed for the high redshift sample. The reason is
that the UV line ratios are not so sensitive to $A_{I/M}$ as the optical
line ratios, which involve lines produced in
partially ionized  regions inside the clouds. The UV lines studied here are produced mainly near the illuminated face of
the emitting clouds.  Unlike at low $z$, where a simple variation
of $A_{I/M}$ can explain the observed trend in the optical line ratios, 
the UV line ratios of the high redshift sample
suggest that  changes in $U$  play
a more important role.

	The harder continuum might be the reason
why the mo\-dels lie closer to the data than the -1.5 PL models (the -1.3 lies between the -1.0 and -1.5 sequences). Adding a matter bounded component
to the previous -1.5 PL mo\-dels will not help to fit the data: CIV/HeII will remain nearly constant and CIII] /HeII
will  decrease slightly  (although overlapping in part with the HeII region,  
CIII]  is formed deeper inside the clouds).

 	Therefore, {\it a mixture of radiation and matter bounded clouds, 
cannot explain the discrepancies between the -1.5 PL models and the observed data}.

\subsection{A diagnostic in the optical}

	Among the possibilities studied here to explain the UV data, low 
 metallicities and a hard power law ionizing continuum produce a good
fit to the data.  As the models overlap in the UV, a way to discriminate between them is to compare the predicted {\it optical} (rest frame) line ratios with the observations.
 One of the few examples of objects with infrared (optical
rest-frame) spectroscopy is the radio galaxy 4C40.36  (Iwamuro {\em et al.} 1996). We have compared 
the optical line ratios with the values derived from 
a) the -1.0 power law models 
and  b) the  -1.5 power law and  0.4 solar metallicity values, both of which
explain the UV line ratios (Table 2).  This object is also included in our sample. The UV line 
ratios (CIV/CIII] =1.05, CIII]/HeII=1.05 and CIV/HeII=1.11) indicate that, for
a hard PL of index $\alpha$=-1, the ionizing parameter U $\sim$0.008 and for the
-1.5 PL and low metallicity models, U $\sim$0.002. The
ratios predicted in the optical are given in Tab.~2: 

\begin{table*}
\centering
\caption{Predicted and observed optical line ratios for the radio galaxy 4C40.36 (the models presented are the ones which reproduce the UV line ratios).}
\begin{tabular}{lllllll} \hline

 & OIII$\lambda$5007/OII$\lambda$3727& OII/H$\beta$	& NeIII$\lambda$3869/H$\beta$ &  HeII$\lambda$4686/H$\beta$  & OIII/H$\beta$ \\ \hline
Observed & 3.3$\pm$1.1 &  5.3$\pm$5.1	&  4.6$\pm$5.6 & 1.0$\pm$1.5 & 17.5$\pm$14.4  
\\ \hline
Models -1.0 PL ($U$=7$\times$10$^{-3}$) & 3.2 & 5.2 & 1.2 & 0.3  & 16.7 \\  
Models low Z ($U\sim$2$\times$10$^{-3}$) & 13.0 & 1.0   &  0.9   & 0.2   & 13.0   & \\
\hline
\end{tabular}
\end{table*}

	The flux of some of the lines is poorly determined and the errors
in the line ratios are large. The most accurate ratio is OIII$\lambda$5007/OII$\lambda$3727 whose value is in very good agreement
with the -1.0 power law model while the low metallicity model can not 
account for it. The high values of the ratios OII/H$\beta$ and OIII/H$\beta$, if confirmed, are also in better agreement with the hard continuum
model.

It is clear  that optical line ratios are able to  discriminate between
these two possibilities.
 The acquisition of good quality optical (rest-frame) spectra of the HZRG
will be crucial to test if the models valid for the UV lines can also fit
the optical line ratios.

\section{Low redshift radio galaxies}

In the above discussion we
used the UV emission lines as a  diagnostic to 
distinguish between
shock and AGN photoionization in the EELR of powerful high redshift radio galaxies.

	At $\it low$ redshifts, most of the information we have comes from
the optical emission lines. The most recent shock models developed by Dopita and Sutherland (1995,1996) produce, in
general, 
optical line ratios which overlap with the -1.5 PL predictions. Therefore, the {\it optical} emission lines do not provide a clear-cut 
way to 
distinguish between the mechanisms.

	However, the UV
 line ratios offer  a powerful diagnostic  for {\it low} redshift radio galaxies
with {\it low} $U$ values (log$U\leq$ -2.0). For {\it high} $U$ values, the distinction is difficult, due to the overlap
of the shock and -1.5 power law models (see above), and the diagnostics become inefficient at the high ionization end of the diagram.
A comparison of Fig.~ 1 and 3 reveals
that the -1.5 power law  with log$U\leq$ -2.0 lie far
from the shock models, due to the much fainter CIV emission. While these AGN
photoionization models produce $log(CIV/CIII] )\leq$0.5 and 
$log(CIV/HeII)\leq$0.2, the
values for the shock models are $\geq$1.0 and $\geq$0.6 and respectively.

	As mentioned before, observations of low redshift radio galaxies with strong jet/cloud interactions
show that the emitting gas has a low ionization level (Clark \& Tadhunter 1996).  The spectrum is dominated by the lines 
emitted by the	``cool'' high density (i.e. low ionization parameter) gas
behind the shock.  Therefore, they
are suitable for the diagnostics. 
A good example is  the radio galaxy  PKS2250-41. Clark
et. al (1996) have presented clear evidence of the presence of 
a jet/cloud interaction in the EELR of
this object both from imaging and optical spectroscopy. 
The optical line ratios
indicate that  log$U\sim$~-3.  Thus the measured
UV line ratios have the potential to determine which mechanism  dominates
the emission line processes in the EELR
of this object.  The possibility of a -1.0 PL ionizing continuum
is rejected because it does not explain the optical line ratios: shock and/or
-1.5 PL (or hot black body) are the main possibilities.

\section{Summary and conclusions}

	We have studied the UV line ratios of a sample of {\it high} redshift
radio galaxies to understand which is the main mechanism responsible for the
line emission processes.

	The  models which reproduce the optical line ratios of low redshift
EELR (-1.5 power law or hot black body and
solar abundances of the gas)
cannot explain the UV line ratios of very high redshift radio galaxies.
We have investigated several possibilities to explain these discrepancies.
The most suggestive one, i.e. shocks produced in jet/cloud interactions, fails to
explain the data.

	On the contrary, a harder ionizing AGN continuum (power law of index $\alpha$=-1) and
the classical -1.5 power law with a low metallicity  produce an
excellent fit to the data. The first possibility is supported by the 
existing evidence of an evolution towards harder continuum at higher redshifts
of the mean spectral shape of the AGN conti\-nuum. The second possibility is
supported by the evidence provided by studies of absorption line systems
in the line of sight of quasars, which demonstrate abundance of $\sim$0.1 solar
at very high redshifts.

The UV diagnostic diagrams do not distinguish unambiguously between
jet-induced shocks and AGN photoioni\-zation  for the HZRG with $high$ ionization state because the two sets of
models overlap in the diagrams.  We note, however,
that not only are the sequences on the diagnostic diagrams
better explained in terms
of photoionization, but the overall ionization state of the HZRG is conside\-rably
higher than is measured in well-studied jet---cloud interactions
at lower redshifts. Our result do not contradict the existence of shocks in these objects.
Rather it suggests that the continuum emitted by the central AGN is the main ionizing source, with the shocks  having an important effect on the kinematics and morphology of the gas.

For the lower ionization states, 
the distinction between the shock and photoionization
models is more clear-cut, and the UV diagnostic diagrams are a
promising means of distinguishing the major ionization mechanism
in the low redshift jet---cloud interaction candidates.

Finally, we note that it is always dangerous to base 
strong conclusions about the
ionization mechanisms on measurements of just a few emission lines in a
particular spectral range. Our results are suggestive but not conclusive.
It is now essential to check the consistency of the model results
by obtaining spectra which cover diagnostics 
in {\it both} the optical and
UV for individual high and low redshift radio galaxies.

\begin{acknowledgements} 
We thank  Luc Binette for his code MAPPINGS I. M.Villar-Mart\'\i n thanks
Jacco van Loon for useful discussions.  M.Villar-Mart\'\i n acknowledges 
support from PPARC grant.
\end{acknowledgements}

\end{document}